# Evidence for monoclinic crystal structure and negative thermal expansion below magnetic transition temperature in Pb(Fe$_{1/2}$Nb$_{1/2}$)O$_3$

Satendra Pal Singh and Dhananjai Pandey[a]

School of Materials Science & Technology, Institute of Technology,

Banaras Hindu University, Varanasi-221 005, India

Songhak Yoon and Sunggi Baik

Department of Materials Science and Engineering,

Pohang University of Science and Technology, Pohang 790-784, Korea

Namsoo Shin

Pohang Accelerator Laboratory,

Pohang University of Science and Technology, Pohang 790-784, Korea

The existing controversy about the room temperature structure of multiferroic Pb(Fe$_{1/2}$Nb$_{1/2}$)O$_3$ is settled using synchrotron powder x-ray diffraction data. Results of Rietveld refinements in the temperature range 300 to 12K reveal that the structure remains monoclinic in the Cm space group down to 12K, but the lattice parameters show anomalies at the magnetic transition temperature (T$_N$) due to spin lattice coupling. The lattice volume exhibits negative thermal expansion behaviour, with $\alpha$ = - 4.64 ×10$^{-6}$ K$^{-1}$, below T$_N$.

[a] Author to whom correspondence should be addressed;
electronic mail: dpandey@bhu.ac.in

Recent years have witnessed enormous interest in multiferroic materials due to their potential applications in memory, sensor and actuator devices[1]. Lead Iron Niobate, $Pb(Fe_{1/2}Nb_{1/2})O_3$ (PFN), is a multiferroic material exhibiting paraelectric to ferroelectric[2] and paramagnetic to G-type antiferromagnetic[3,4] transitions at 385K and 143K, respectively. It is an attractive material for use in multilayer ceramic capacitors and other electronic devices due to its high dielectric constant (>10,000), diffuse phase transition behaviour[5] and low sintering temperature[6]. However, there exist controversies about the structure of PFN at room temperature until now. Both rhombohedral[7] and monoclinic[8,9] structures in the R3m and Cm space groups, respectively, have been proposed, but more careful investigation is still needed to obtain a clear picture about the correct crystal structure of PFN. In a recent temperature dependent dielectric study, a jump in the dielectric constant near the Neel temperature $T_N$ = 143K has been reported[4]. Using earlier Landau theory results[10], Yang et al. have shown that the change in the dielectric constant at Neel temperature may be associated with the magnetoelectric coupling term $\gamma P^2 M^2$, where P, M and $\gamma$ are polarization, magnetization and magnetoelectric coupling coefficient, respectively. A similar change in the dielectric constant at Neel temperature has been reported in other magnetoelectric materials such as $RMn_2O_5$ and $RMnO_3$ (R = Tb, Ho, Dy)[1,10-13]. In the $RMn_2O_5$ family of magnetoelectrics, changes in the cell parameters at the Neel temperature have also been reported and it has been interpreted as a signature of spin-lattice coupling in such magnetoelectrics[12,13]. No attempt has been made so far to look for the anomalies in the unit cell parameters as a result of magnetoelectric coupling in PFN. In this letter, we present the results of Rietveld analysis of high resolution synchrotron x-ray diffraction (XRD) data to resolve the existing controversies about the structure of PFN at room temperature. After settling the room temperature structure of PFN, we have also carried out Rietveld analysis of powder XRD data, collected at different temperatures in the range 300 to 12K, to see if there is any lattice parameter anomaly associated with the magnetic transition. It is shown that the structure of PFN is monoclinic in the Cm space group in the entire temperature range 300-12K of our study. It is also shown that the lattice parameters and unit cell volume show distinct anomaly at $T_N$, with unambiguous evidence of negative thermal expansion below $T_N$.

Pyrochlore free PFN samples were prepared by solid-state route, the details of which are described elsewhere[14]. For x-ray characterization, the sintered pellets were crushed to fine powders and then annealed at 500°C for 10 hours to remove the strains introduced during crushing. X-ray diffraction (XRD) measurements were carried out using an 18 kW rotating anode (Cu $K_\alpha$) based Rigaku powder diffractometer operating in the Bragg- Brentano geometry and fitted with a graphite monochromator in the diffracted beam and attached with a close cycle He-refrigerator for varying the sample temperature continuously in the temperature range 300 to 12K. The data was collected in the 2θ range of 20 to 120 degrees at a step length of 0.02 degrees during heating after cooling the sample to 12K. Synchrotron powder XRD experiments were carried out at 8C2 HRPD beamline at Pohang Light Source (PLS). The incident x-rays were monochromatized to the wavelength of 1.543 Å by a double bounce Si (111) monochromator. The diffraction pattern was scanned in the 2θ range of 20 to 130 degrees at a step length of 0.01 degrees. Rietveld refinements were carried out using Fullprof program[15]. In the refinements, pseudo-Voigt function and a fifth order polynomial were used to define the profile shape and the background, respectively. Except for the occupancy parameters of the ions, which were fixed at the nominal composition, all other parameters, such as scale factor, zero correction, background, half-width parameters, the mixing parameters, lattice parameters, positional coordinates, and thermal parameters, were varied in the course of refinement. It was found necessary to use anisotropic peak broadening for the synchrotron data, whereas the laboratory data could be analysed using isotropic peak broadening function only. The isotropic thermal parameter for Pb was found to be considerably large (~ 2.216) indicating Pb-site disorder, as reported by earlier workers[9] and the use of anisotropic thermal parameters in the refinements resulted in lower $\chi^2$ values. For the rhombohedral phase with R3m space group, we have used hexagonal axes with lattice parameters $a_H = b_H = \sqrt{2} a_R$ and $c_H = \sqrt{3} a_R$, where $a_R$ corresponds to the rhombohedral cell parameter.

Synthesis of phase pure perovskite PFN is a major challenge, as the pyrochlore phases such as $Pb_2Nb_2O_7$ and $Pb_2Nb_4O_{13}$ get easily formed[16]. Using a modified solid state route[14], we were able to synthesize pyrochlore free PFN samples. Figure 1 depicts the synchrotron powder XRD pattern of PFN at room temperature. There are no peaks near $2\theta \approx 28.84$ or 29.25 degrees, which are the strongest XRD peaks for the $Pb_2Nb_2O_7$ and $Pb_2Nb_4O_{13}$, confirming the absence of the pyrochlore phase. All the peaks in this figure correspond to the perovskite PFN phase. The inset depicts the zoomed profiles of the pseudocubic 200 and 222 reflections which are singlet and doublet, respectively, indicating a rhombohedral structure at first sight. However, if 200 peak is truly singlet, as expected for the rhombohedral structure, its width should have been less than that of the pseudocubic 222 peak following Caglioti relationship for the two-theta dependence of the peak width[17]. The width of 200 peak is about 1.6 times that of the 222 pseudocubic peak and it suggests that 200 peak is not singlet and hence the true structure may not be rhombohedral. A similar anomalous broadening of the 200 peak in $Pb(Zr_xTi_{1-x})O_3$ (PZT) and (1-

x)[Pb(Mg$_{1/3}$Nb$_{2/3}$)O$_3$]-xPbTiO$_3$ (PMN-xPT) has been attributed to a short range ordered monoclinic phase in the Cm space group[18]. A choice between the R3m and Cm space groups in these materials was made unambiguously using profile refinement techniques.

Figure 2 presents the Rietveld fits for 200, 220 and 222 pseudocubic profiles obtained after full pattern refinements using rhombohedral and monoclinic structural models for PFN. For the rhombohedral model, if we try to account for the large broadening of the 200 reflection, the fit for the 222 reflection becomes very poor as is evident from Fig. 2(a). On the other hand, if we try to force good fit for the 222 reflection, the fit for 200 and other reflections becomes poor, as shown in Fig. 2(b). Very good fit, however, is obtained for all the other reflections if one uses the monoclinic structure in the Cm space group as shown in Fig. 2(c). The agreement factors, the DW statistics[19] and the Prince's criterion[14] favor the Cm space group. Table 1 lists the refined structural parameters for the monoclinic Cm space group. The equivalent elementary perovskite cell parameters of the monoclinic Cm of PFN bear the relationship $a_m/\sqrt{2} \approx b_m/\sqrt{2} < c_m$, and hence this phase is of M$_A$ type in the notation of Vanderbilt and Cohen[20].

Having settled the room temperature structure of PFN unambiguously, we carried out Rietveld refinements with the monoclinic structure in the Cm space group for the low temperature XRD data in order to investigate the effect of magnetoelectric coupling on the structure of PFN near T$_N$. We find that the monoclinic structure remains unchanged below T$_N$. Figure 3 shows lattice parameters, unit cell volume and the monoclinic distortion angle (β) of PFN as a function of temperature. All the three lattice parameters (a, b and c) show anomalies around 150K, which is close to the magnetic transition temperature reported in the literature[21]. The monoclinic distortion angle (β), however, does not show any anomaly and increases continuously as the temperature decreases. The lattice parameter 'b' becomes nearly temperature independent below 150K whereas lattice parameters 'a' and 'c' exhibit negative thermal expansion. The volume of the unit cell first decreases on cooling up to 150K and then starts to increase below 150K, showing a negative volume thermal expansion at T< $T_N$. The linear negative thermal expansion (NTE) coefficient (α) obtained from the fit in Fig. 3 (b) at T ≤ 150K is -4.64 ×$10^{-6}$ $K^{-1}$.

The observation of anomalies in the temperature dependence of dielectric constant[10,11] and ferroelectric polarization at magnetic transition temperatures are taken as evidence for magnetoelectric effect due to spin lattice coupling[11,22]. Such a coupling is also expected to lead to anomalies in the lattice parameters. However, the earlier studies for resolving such lattice parameter anomalies using scattering techniques in materials like $HoMn_2O_5$ and $DyMn_2O_5$[13] have failed to provide any evidence. Very weak anomalies at $T_N$ have been reported in $TbMn_2O_5$[12]. In $DyMn_2O_5$, this anomaly is somewhat more pronounced and there is a sign of negative thermal expansion also below $T_N$[13]. In comparison, the lattice parameter and the unit cell volume anomalies are well pronounced in PFN. The existence of negative thermal expansion below $T_N$ clearly suggests that the thermal contraction below $T_N$ due to anharmonicity is being more than offset by the magnetic ordering. We believe that the ferromagnetic component of the G-type antiferromagnetic state below $T_N$ is responsible for the negative thermal expansion due to the spin-lattice coupling. In $SrRuO_3$, there is also competition between the lattice and the magnetic contributions to the overall thermal expansion behaviour below the magnetic transition temperature[23]. This has been attributed to a magnetovolume effect arising from itinerant electron magnetism. It remains to be seen whether this mechanism is responsible for the negative thermal expansion in PFN. We hope that our results will encourage some *ab initio* first principle calculations to understand the role of magnetic ordering on the thermal expansion behaviour, which may in turn throw light on the magnetoelectric coupling.

Satendra Pal Singh acknowledges financial support from the All India Council of Technical Education (AICTE) in the form of the award of a National Doctoral Fellowship (NDF). The experiments at Pohang Light Source (PLS) were supported by the Ministry of Science and Technology (MOST) and POSTECH, Pohang, Korea.

Table I

Refined structural parameters of Pb(Fe$_{1/2}$Nb$_{1/2}$)O$_3$ using monoclinic structure in the Cm space group.

$a_m$ = 5.6787(1) Å; $b_m$ = 5.67310(9) Å; $c_m$ = 4.01520(9) Å;
$\alpha = \gamma$ = 90.00 and $\beta$= 90.098(7) (degrees)

| Ions | x | y | z | B(Å$^2$) |
|---|---|---|---|---|
| Pb$^{+2}$ | 0.0000 | 0.0000 | 0.0000 | $\beta_{11}$ = 0.015(2) |
|  |  |  |  | $\beta_{22}$ = 0.022(2) |
|  |  |  |  | $\beta_{33}$ = 0.029(4) |
|  |  |  |  | $\beta_{13}$ = 0.008(1) |
| Fe$^{+3}$/Nb$^{+5}$ | 0.510(3) | 0.0000 | 0.478(3) | B= 0.23(7) |
| O$^{-2}_{I}$ | 0.53(1) | 0.0000 | -0.04(1) | B= 0.6(3) |
| O$^{-2}_{II}$ | 0.273(8) | 0.254(8) | 0.44(1) | B= 0.4(2) |

$R_p$ = 8.69; $R_{wp}$ = 11.9; $R_{exp}$ = 9.53; $\chi^2$ = 1.56

**Figure captions:**

Fig. 1. Synchrotron x-ray powder diffraction pattern of Pb(Fe$_{1/2}$Nb$_{1/2}$)O$_3$ at room temperature. Insets (a) and (b) show the zoomed profiles of the pseudocubic 200 and 222 reflections, respectively.

Fig. 2. Observed (dots), calculated (continuous line), and difference (bottom line) profiles of the 200, 220 and 222 pseudocubic reflections obtained after full pattern Rietveld refinements using the room temperature synchrotron powder diffraction data of Pb(Fe$_{1/2}$Nb$_{1/2}$)O$_3$ in the 2θ range 20 to 130 degrees: (a) and (b) rhombohedral R3m space group and (c) monoclinic Cm space group. The tick marks above the difference plot show the position of the Bragg peaks.

Fig. 3. Temperature dependent variation of (a) lattice parameters (a, b and c), (b) unit cell volume and (c) the monoclinic distortion angle (β) obtained from Rietveld refinements using powder x-ray diffraction data. The equivalent elementary perovskite cell parameters are calculated as a = a$_m$/√2, b = a$_m$/√2 and c = c$_m$.

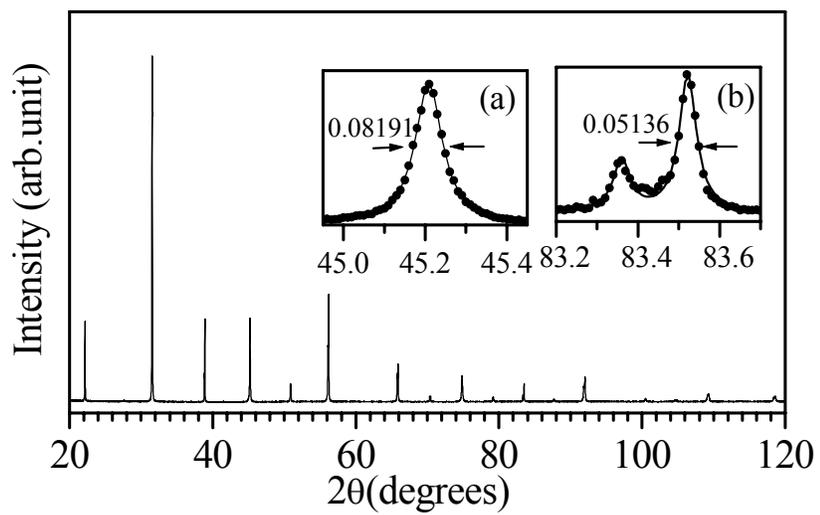

Fig. 1

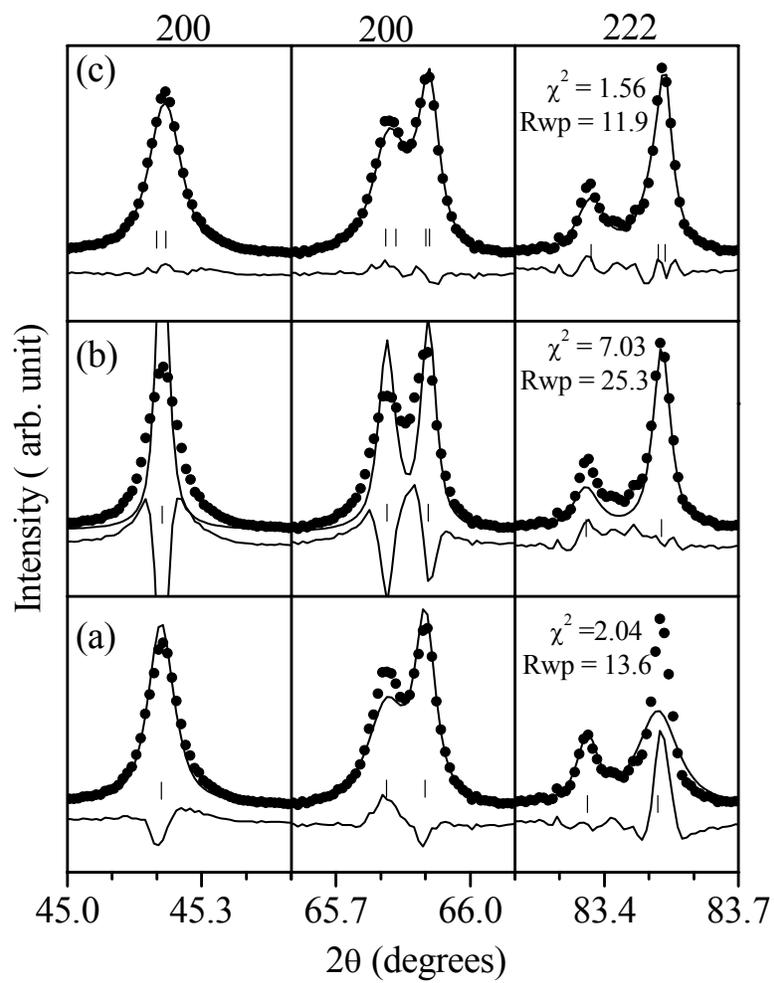

Fig. 2

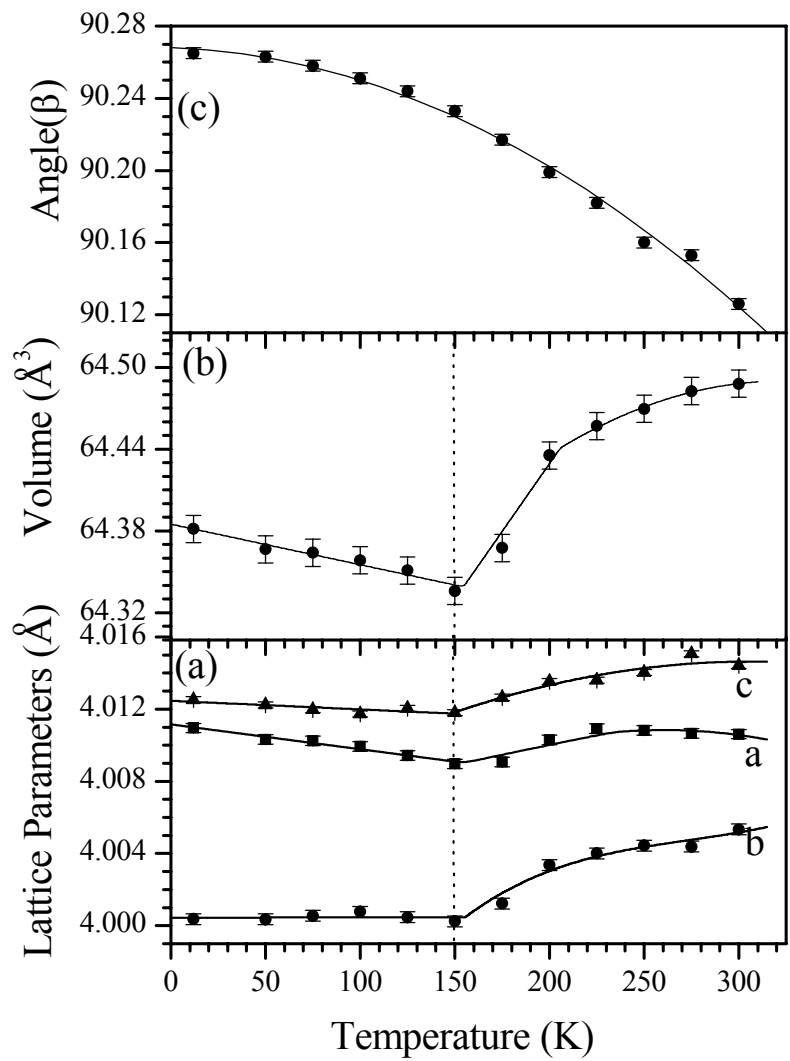

Fig. 3